\documentclass[aps,prb,preprint,groupedaddress]{revtex4}
\usepackage{graphicx}

\begin{document}


\title{Quantum entanglement, fair sampling, and reality:\\Is the moon there when nobody looks?}


\author{Guillaume ADENIER}
\affiliation{International Center for Mathematical Modeling in
Physics and Cognitive Sciences, V\"{a}xj\"{o} University, 35 195
V\"{a}xj\"{o}, Sweden}


\date{\today}

\begin{abstract}
In 1981, David Mermin \cite{Mermin1981a,Mermin1981b} described a
cleverly simplified version of Bell's theorem \cite{bell64}. It
pointed out in a straightforward way that interpreting
entanglement from a local realist point of view can be
problematic. I propose here an extended version of Mermin's
device that can actually be given a simple local realist
interpretation through a sample selection bias, and I argue that we still have no scientific
reason to believe that the moon could possibly not be there when
nobody looks \cite{Mermin1981a}.
\end{abstract}

\pacs{}

\maketitle


Entanglement is arguably the greatest mystery brought forth by
Quantum mechanics: it has such drastic consequences on the
possible physical models that can explain its features that the
philosophical representation of the world itself is at stake.

Describing the mystery of entanglement can be difficult. It is
sometimes misrepresented as a miraculously perfect correlation
between two distant objects. This picture is however insufficient,
as the same perfect correlation can be obtained with a simple
local realistic model---given by Bell himself
\cite{bell64}---which would not violate any Bell inequality. It is
only when one deviates from identical measurements that the
strangeness of entanglement reveals itself. David Mermin
illustrated this in a simple version of Bell's theorem that
stripped the mystery of entanglement to the core.

Let us start first by recalling the way Mermin's device works. It
consists of three unconnected black boxes. The first two boxes,
labelled A and B, are detectors. Each detector has a switch with
three possible positions (1, 2, or 3), and two lamps (a green one
labelled G, and a red one labelled R). Whenever a particle reaches
a detector, one of the two lamps fires, and a measurement result
is recorded accordingly (G or R).

\begin{figure}
\center
\includegraphics[width=10cm]{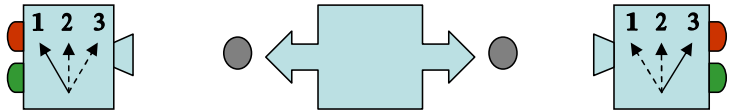}
\caption{\label{fig:moon0} Mermin's device. A and B are detectors
with random switches, C is a source of pairs of particles sent to
the detectors.}
\end{figure}

The third black box, labelled C, is the source: it sends pairs of
particles going in opposite directions: the first particle going
to detector A, the second going to detector B. The detectors can
be arbitrarily far from each other and are not connected in any
way, so that the detectors have no influence upon one another.

The switches on detectors A and B are randomly and independently
selected after a pair of particles has left the source and before
either particle has arrived at its detector. There are thus nine
equally probable settings for the pair of detectors: 11, 12, 13,
21, 22, 23, 31, 32, and 33 (the first and second digits represent
the position of the switch for detectors A and B respectively).
There are similarly four possible pairs of measurement results:
RR, RG, GR, and GG. A measurement of a pair of particles is thus a
quadruple of two digits and two letters (like 21GR for instance).

The source then fires an arbitrary large number of pairs and the
behaviour of Mermin's device can be summarized in the following
way:

\begin{itemize}
    \item \emph{Case a}. In those runs in which the detectors have the same
settings (11, 22, or 33), the lights always flash the same colour,
RR and GG appearing with an equal frequency.
    \item \emph{Case b}. In those
runs in which the detectors have different settings (12, 13, 21,
23, 31, 32), the lights flash the same colour with a frequency of
only 1/4.
\end{itemize}

The challenge proposed by Mermin is to provide a model that can
explain both \emph{Case a} and \emph{Case b}, given that the
detectors A and B have no known connection between each other.

The perfect correlation observed in \emph{Case a} is a strong
constraint on all possible models. The only simple way to explain
this perfect correlation is to assume that each particle carries
instructions specifying which colour will flash for each of the
three possible settings. Furthermore that particles belonging to
the same pair carry identical instructions. Instruction sets must
provide an instruction for each of the three possible settings 1,
2, and 3, because the settings are not chosen until after the
particles have separated, and each of the pairs of settings 11,
22, and 33 has 1/9 of a chance of being chosen in any given run.
There are eight possible instruction sets given to each particle:
RRR, RRG, RGR, RGG, GRR, GRG, GGR and GGG. Mermin shows that
fulfilling \emph{Case a} with such instruction sets introduces an
excess of ``same colour" results in \emph{Case b}. In order to see
this, let us consider the instruction set GGR. This instruction
set means that a particle carrying it will flash G for settings 1
and 2, and R for setting 3. In \emph{Case a} the same measurement
results are observed on both sides since the particles from one
pair carry the same instructions (see Fig. 2).

\begin{figure}
\center
\includegraphics[width=5cm]{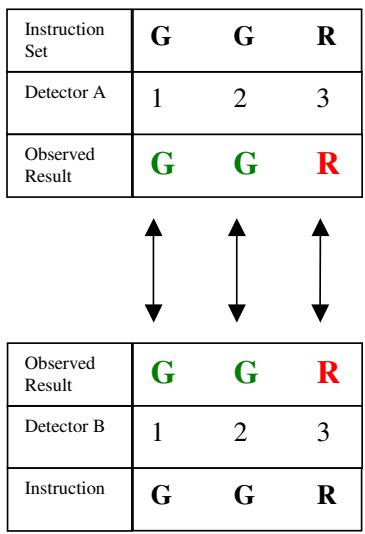}
\caption{\label{fig:moon1} \emph{Case a} (same settings) always
result in the same colour result (GG or RR).}
\end{figure}

Let us now consider \emph{Case b}, that is, when the switches A
and B are set on different positions, which is graphically
represented as diagonal lines. There are six possible ways to draw
such diagonal lines, four short ones (see Fig. 3a), and two long
ones (see Fig. 3b). There are two ways to obtain the same
measurement results with diagonal lines, that is 12GG and 21GG.
The remaining four possible diagonals give opposite results, that
is, 13GR, 31RG, 23GR, and 32RG.

\begin{figure}
(a)
\includegraphics[width=5cm]{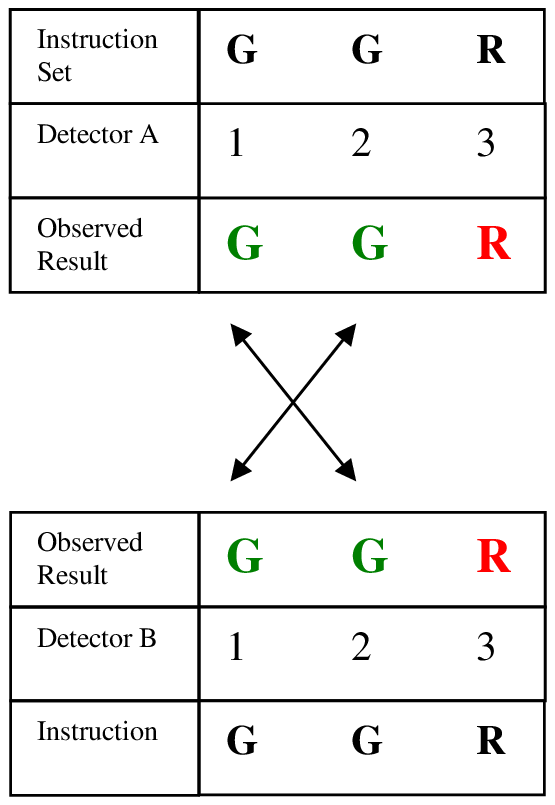}
(b)
\includegraphics[width=5cm]{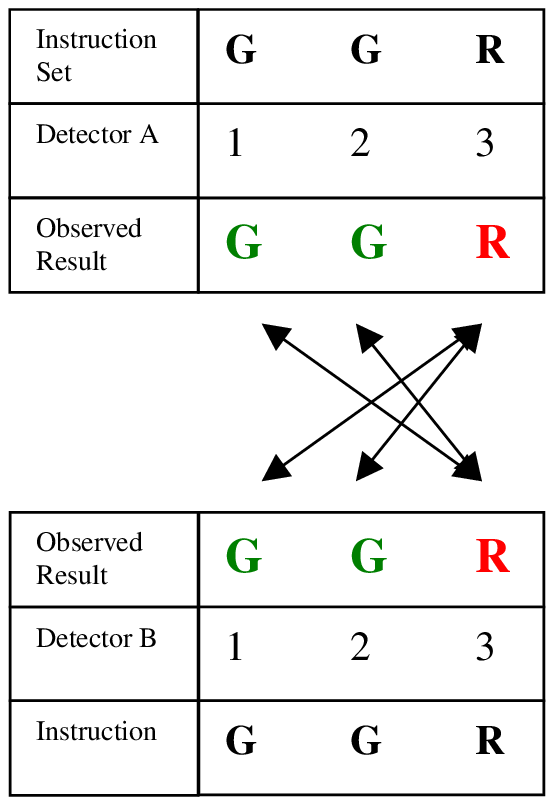}
\caption{\label{fig:moon2} \emph{Case b} (different settings)
resulting in (a) same colour results, or (b) opposite colour.}
\end{figure}

Hence the totals are two out of six pairs of switch positions with
the same colour results, and four out of six with opposite colour
results. Therefore, if the source was sending only instruction set
GGR, the proportion of same colour in \emph{Case b} would be 1/3,
and the proportion of opposite colour results would be 2/3.

The same reasoning for \emph{Case b} applies to any of the
instruction sets RRG, RGR, RGG, GRR, and GRG, since they all share
with GGR that one colour appears in the instruction set once and
the other colour twice. Hence, if the source is sending any of
theses 6 possible states, the \emph{Case b} proportion of same
colour results should be 1/3, and the proportion of opposite
colour results should be 2/3 (independently of the specific
probability distribution of the states).

The two remaining instruction sets, RRR and GGG, always give the
same colour results, independently of the switch settings. If we
thus want to minimize the amount of same results in \emph{Case b}
(while keeping the \emph{Case a} feature), the source C should
never send any of the homogeneous states RRR and GGG. Even so, in
the long run the observed frequency of same results in \emph{Case
b} should be close to 1/3, which is significantly higher than the
1/4 specified by Mermin's device. This leaves us with a conundrum,
as the only reasonable way to account for \emph{Case a} is
incompatible with an account for \emph{Case b}.

As stressed by Mermin, experiments done since Bell's paper are in
agreement with quantum-theoretic predictions, so that Mermin's
device could in principle be constructed. However, this device would
not behave entirely according to Mermin's expectation. Indeed, all
experiments performed so far share a common feature: most particles sent to the detectors in order to
exhibit the conundrum remain
undetected\cite{Aspect82,AspectTh,Weihs}.

The fact that most particles remain undetected implies that no
direct and complete information on the population of all emitted
particles is possible. The issue that it raises is not specific to
experiments on quantum entanglement. It is quite general in
statistical analysis as soon as obtaining data from the entire
population under study proves impossible. The statistics are then
restricted to a sample of that population. In order to make any
inference on the target population, one has to make sure that no
selection bias enters in the sampling procedure: the sample must
fairly represent the population.

To start with, the selection method itself may introduce a bias. Take for
example a survey research that for practicality is only selecting households with landline
telephones. It will automatically exclude individuals having no telephone
or only mobile phones, which represents a non negligible and distinctive part of the population.

A selection bias may also arise due to the type of question asked in
the survey, inducing some self selection of the individuals in the
sample (non-response error). Exemplifying this would be a survey on the negative
or positive perception of surveys in general. It would be likely to be
biased favorably towards surveys, simply because individuals with a
negative opinion would be more likely to refuse to participate in
the survey.

In the context of EPR-Bell experiments, the possibility that a
sample selection bias could account for the observed correlations
was raised first in 1970 by Pearle\cite{Pearle}, and has been
discussed many times since
\cite{ClauserS,Garra,GargMermin,Santos,Gisin,Larsson1,Larsson2,PappaR,thompson}.
It is often referred to as the ``detection efficiency loophole", but
it is however a rather misleading terminology, as it already
contains an implicit interpretational choice: the blame is
implicitly put on the detectors imperfections. As we will see, this
interpretation leaves the conundrum intact. Another interpretation
of these non-detections in terms of a selection bias is however
possible, and we will see that it can resolve the
conundrum\footnote{After this paper was submitted, a referee pointed
out that the idea of a sample selection bias in the context of
Mermin's device had been made by Mermin himself (on page 168 of his
collection of articles\cite{Boojums}), as well as in the Letters
section of Physics Today\cite{PhysTodayNov85} (commenting on another
version\cite{PhysTodayAp85} of Mermin's
demonstration\cite{Mermin1981a,Mermin1981b}) both in two letters
--- one by Marshall and Santos, the other by Jordan --- and in a commentary on those letters, in which Mermin
points out that any correlations whatever can be produced by having
two such ``don't fire" instruction in each instruction set.}.

We thus consider an extended version of Mermin's device, in which
whenever a particle enters a detector, it can either induce a green
flash (G), a red flash (R), or no flash at all (N). In addition to
the four possible flash results (RR, RG, GR and GG), we then have
four possible results inducing only one flash (RN, NR, GN, and NG),
and one with no flash at all (NN). Whenever no flash is recorded (N)
the whole pair is simply discarded from the statistics since no
flash correlation can be established. This might seem a rather
questionable way to deal with no-flash events, but this is precisely
the way non-detected pairs are treated in real experiments, as the
statistics are computed exclusively on the sub-ensemble of pairs
actually detected.

There are basically two simple ways to account for the occurrence of
no-flash events in the framework of Mermin's device: they can be
either carried by the detectors (independently of the incoming
instruction set), or by the particles themselves.

Let us first consider the case where the absence of flash is due to
the detectors. It means that for some reason the detectors are not
very efficient and that they fail to fire now and then when they
should have, independently of the incoming instruction set. Whatever
causes this unreliability of the detectors, it can be symbolize by
adding an additional possible switch position, that we conveniently
label 0, and by stating that the position of the switch is now
randomly chosen between those four positions, instead of the three
positions in Mermin's original device (see Fig. 4). As a matter of
detail, let p be the probability that the switch is on position 0
then the probability associated to any of the remaining three
switches is equal to (1-p)/3. Whenever the switch turns out to be
set on this additional position (0), none of the lamps can flash,
whatever the incoming particle.

\begin{figure}
\center
\includegraphics[width=5cm]{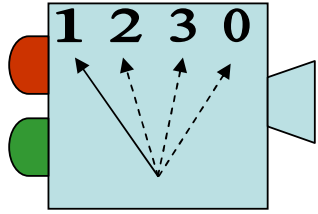}
\caption{\label{fig:moon2bis} No-flash events carried by a fourth
switch (labelled 0) in the detectors.}
\end{figure}

A look at Fig. 5 representing this new situation shows that
Mermin's line of thought, as used above, is still valid. The
no-flash events N always occur on the additional column
corresponding to the failure of the detector, but the statistics
on the pairs producing a flash concerns only the first three
columns, and thus remain the same as obtained previously. Hence,
with this additional switch position, the conundrum remains
intact.

\begin{figure}
(a)
\includegraphics[width=5cm]{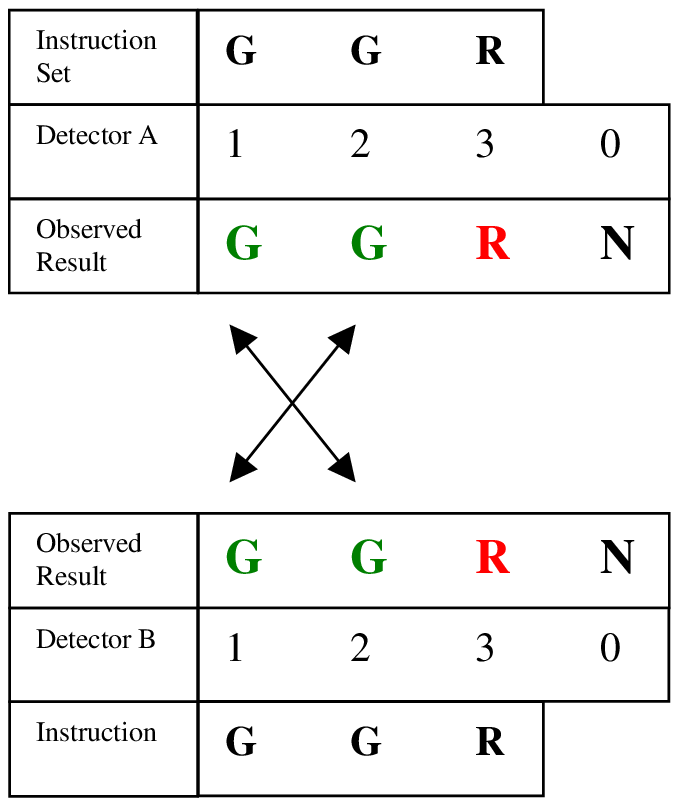}
(b)
\includegraphics[width=5cm]{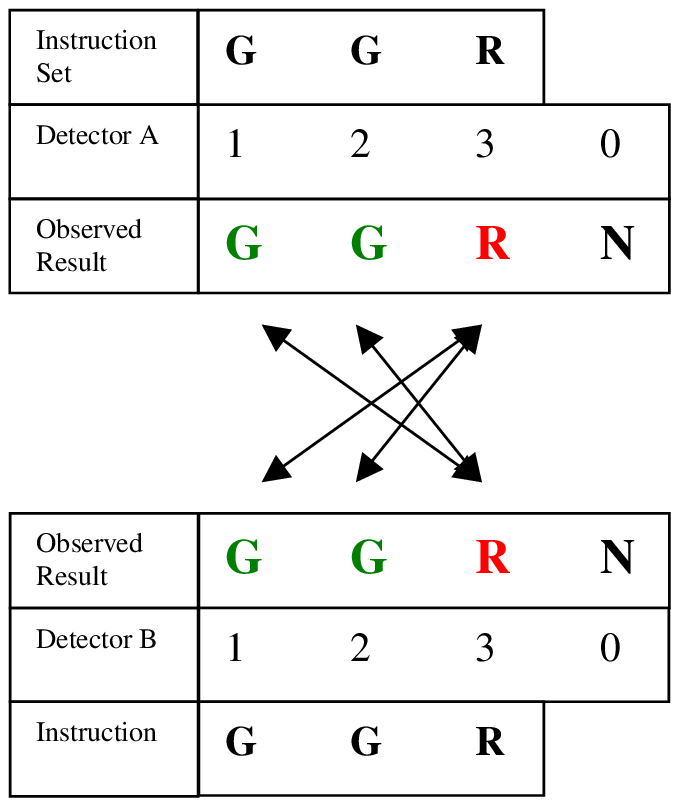}
\caption{\label{fig:moon3} Non-detection with carried by the
detectors. Any result with a no-flash event (N) is discarded. The
statistics of the observed results remain the same, as they
concern only the first three columns. The possible results of
\emph{Case b} are displayed here, and remain the same as in Fig.
\ref{fig:moon2}. More runs are necessary to achieve the same
statistical relevance, but the conundrum is intact. }
\end{figure}

If however the no-flash events are carried by the particles
themselves, the conundrum can be resolved. Consider for instance
the same GGR pair we considered above, but this time with the
freedom to put an instruction of the type ``do not flash", that we
label N, at a specific place in the instruction sets. We know that
for this particular state we have an excess of the diagonal GG
correlations (same colour results, in \emph{Case b}). Let us then
put a N in place of one of the G instruction for the first
particles, say the G corresponding the switch 2, so that the
particle going to A has the instruction GNR. The second particle,
the one going to detector B, carries the initial GGR instruction
set. Since the particles from one pair no longer carry the same
instruction, we now use an explicit notation for each particle,
and we denote for instance this particular pair state by GNR-GGR,
meaning that the particle going to detector A has instruction set
GNR, while the second has instruction set GGR.

Let us first consider \emph{Case a}. As can be seen from Fig. 6,
it still gives a perfect correlation for the detected pairs, since
the results are the same colour in two cases: (11GG, 33RR), while
the third is discarded because detector A does not flash (22NG).

\begin{figure}
\center
\includegraphics[width=5cm]{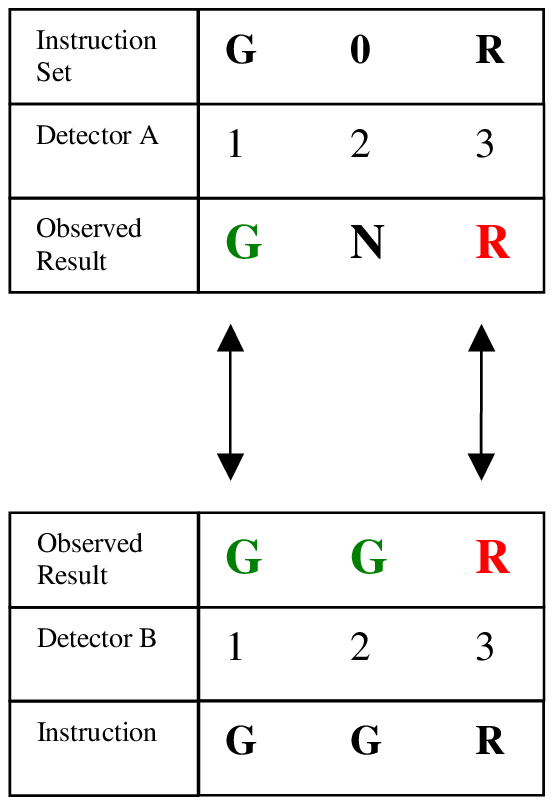}
\caption{\label{fig:moon4} Non-detection carried by the
instruction sets in \emph{Case a}. Any result with a no-flash
event (N) is discarded.}
\end{figure}

If we now consider \emph{Case b}, we see (Fig. 7 a) that we retain
only one possibility to obtain the same colour result  (12GG),
while the other one is discarded because the particle entering in
detector A produces no flash (21NG). Similarly, there are three
possibilities (Fig. 7 b) to obtain opposite colour results (13GR,
31RG, 32RG), while the remaining one is discarded because the
particle entering in detector A produces no flash (23NR).

\begin{figure}
\center (a)
\includegraphics[width=4.5cm]{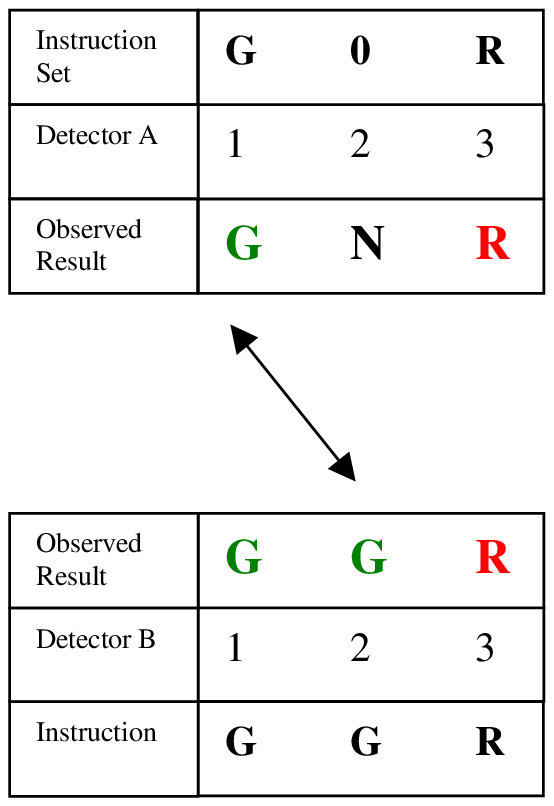}
(b)
\includegraphics[width=4.5cm]{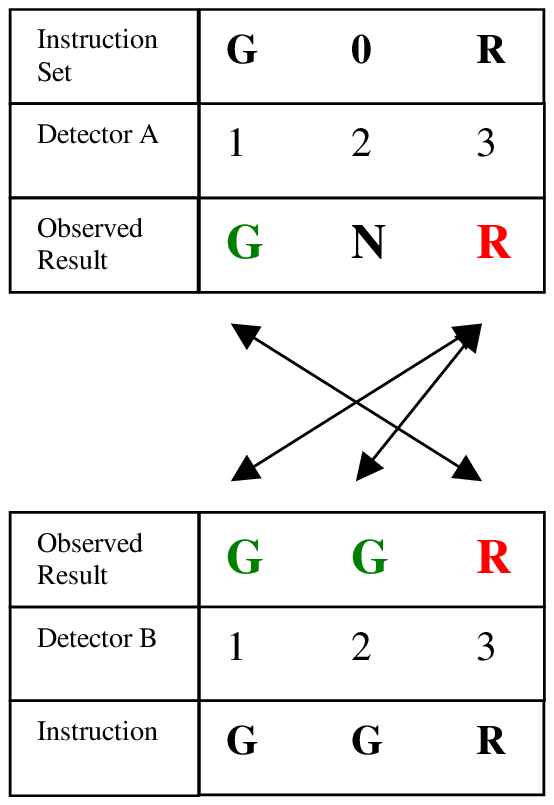}
\caption{\label{fig:moon5} Non-detection carried by the
instruction sets in \emph{Case b}. Any result with a no-flash
event (N) is discarded. }
\end{figure}

Thus, out of the 4 possible ways to register a \emph{Case b} with
the GNR-GGR state, we then have one possibility with same colour,
and three possibilities with opposite colours. In other words, if
the source was sending only this state, the lights would flash the
same colour with a frequency of 1/4 in \emph{Case b}, which is
exactly the specified frequency of Mermin's device.

We can use the same logic for any of the remaining instruction
sets RRG, RGR, RGG, GRR, and GRG. All we need to do is put a
no-flash instruction N in place of one of the colour that appear
twice in the instruction set, either for the particle going to
detector A or for the one going to detector B. The list of
instruction sets displayed on Table I was built following this
idea (the last six instruction sets are the same as the first six
instruction sets after particle exchange):

\begin{table}\label{tablemoon}
\center
\begin{tabular}{|c|c|}
  \hline
  Alice & Bob \\
  \hline
  NRG & GRG \\
  NGR & RGR \\
  RNG & RRG \\
  GNR & GGR \\
  RGN & RGG \\
  GRN & GRR \\
  GRG & NRG \\
  RGR & NGR \\
  RRG & RNG \\
  GGR & GNR \\
  RGG & RGN \\
  GRR & GRN \\
  \hline
\end{tabular}
  \caption{A list of instruction sets satisfying both \emph{Case a} and \emph{Case b} specifications.}
\end{table}

Hence, if the source sends a uniform distribution of these twelve
states, the proportion of same colour results in \emph{Case b}
will therefore be close to 1/4 in the long run, and the proportion
of opposite colour results close to 3/4, while keeping at the same
time the perfect correlation of \emph{Case a}. The conundrum is
resolved.

Note that each switch position (each column) will in the long run
receive the same number of instructions N, G and R, so that the
detectors are equivalent and balanced. The three switches of a
detector and the twelve possible states being random and
independent, there are thirty-six possible combinations of both,
among which only six no-flash instructions N per detector, so a
detector will flash 5/6 of the time (circa 83\%). Interesting
enough, this efficiency is very similar to the known bound on
detectors efficiency required to validate a genuine violation of
Bell inequalities \cite{GargMermin,Larsson1}.

In all real EPR-Bell experiments meant to challenge local realism,
the detectors had a much lower rate of detection. This can be
accounted for in this context by considering mixed process in which
not only the detectors have a fourth switch (0) yielding no-flash,
but the instruction sets carry no-flash instructions (N) as well
(see Fig. 8). This allows in principle to retain the resolution of
Mermin's conundrum even in the case of detectors $A$ and $B$ having
distinct efficiencies $\eta^A$ and $\eta^B$. One needs only to
distinguish the fair sampling part $\eta_f$ in the probability of a
flash from the unfair sampling part $\eta_u$. The former corresponds
to the possibility that the detector itself may fail to fire when it
should have, independently of the state of the incoming particle,
whereas the later corresponds to the possibility that the incoming
particle may have a no-flash instruction for the selected
measurement setting. The probability $\eta$ that a detector fires is
thus the product of these two independent probabilities, that is
$\eta=\eta_u \eta_f$. In order to keep the resolution of Mermin's
conundrum, each particle needs to carry only one single instruction
set from the list given in Table I, and the unfair sampling
efficiency $\eta_u$ should thus remain equal to 5/6. In order to
modelize two detectors with different efficiencies, one needs only
to consider detectors with two different fair sampling efficiencies
$\eta^A_f$ and $\eta^B_f$, so that the corresponding efficiencies of
these detectors are $\eta^A=\eta_u \eta^A_f$ and $\eta^B=\eta_u
\eta^B_f$. Having detectors with different $\eta_f$ does not change
the correlation since the sampling process associated to it is fair:
it just reduces the number of pairs detected.

\begin{figure}
\center
\includegraphics[width=5cm]{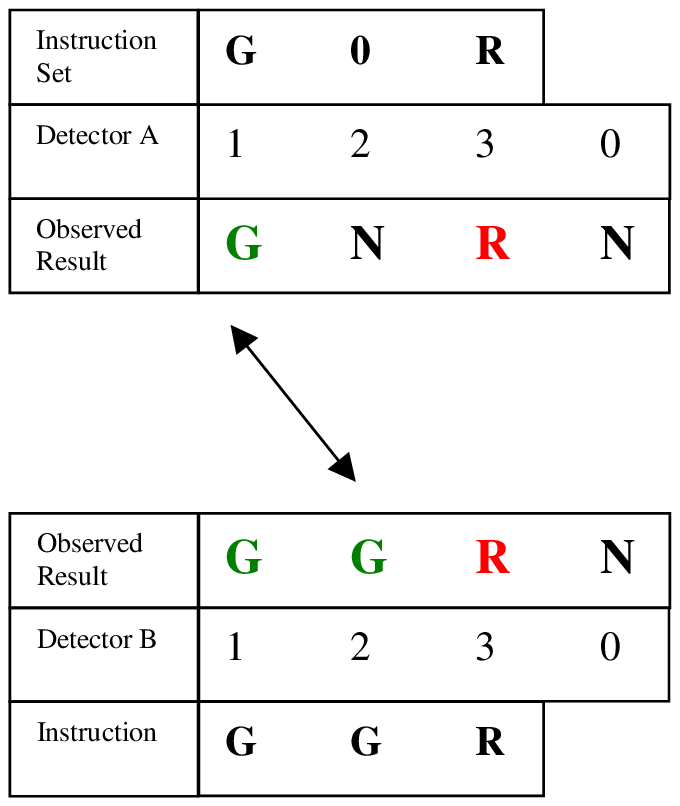}
\caption{\label{fig:moon6} Non-detection carried by both the
detectors and the instruction sets. The proportion of rejected
particles can be arbitrarily large (at least 1/6) and thus
reproduce the statistics of a real implementation of Mermin's
device.}
\end{figure}

Our extended version of Mermin's device shows that the proof of the
conundrum rely on the validity of an extra assumption. The
alternative can be summarized as follows: either the instruction
sets carry no-flash instructions and the conundrum is resolved, or
they do not, and the conundrum holds. The assumption that the second
alternative holds is known in EPR-Bell experiments as the fair
sampling assumption. It is tempting to validate this assumption on a
theoretical basis, arguing that Quantum mechanics does not predict
that the sampling would be unfair (a feature that can however be
accounted for in the framework of contextual probability theory
\cite{KHR10}), but since it is precisely the completeness of Quantum
Mechanics that is at stake in EPR-Bell experiments, this would
render the argument circular, and thus invalid. Consequently, the
fair sampling assumption must be justified empirically. There is
however little experimental evidence to support the validity of this
assumption. One experimental feature that can be brought forward in
support of fair sampling is that the observed size of the detected
sample pairs is independent of the measurement settings. This
feature is however shared by non-symmetrical models like Larsson's
\cite{Larsson1}, and our extended version of Mermin's device shares
this feature as well. As shown elsewhere \cite{AdenierKhrennikov},
the validity of the fair sampling can actually be questioned on the
basis of experimental data.

A detailed discussion about photon detection issues and about the
possible constraints that a sample selection bias would impose on
real detectors is beyond the scope of this article. A valid concern
is that the existence of such a selection bias should have been
spotted in optical experiments outside the EPR-Bell domain. It
should however be noted that two-photon sources used in EPR-Bell
experiments are special in other ways than the alleged quantum
nonlocality. They can for instance be seen as a conditional source
of single photons\cite{Pittman}, so that it is conceivable that such
selection bias would be elusive as it would be a characteristic of
single photon sources exclusively. It is clear that this should be
experimentally testable, and other measurement systems that could
detect such a selection bias have already been
proposed\cite{AdenierKhrennikovTest}. Naturally, this would not be
necessary if high efficient EPR-Bell experiments were within reach,
but as long as this is not the case, and given that such experiments
might be impossible in actuality \cite{Santos2,Percival}, the
possibility of a selection bias should be investigated through both
theoretical and experimental means.

In the meantime, analysis shows that it indeed is still possible to
ascribe properties to objects independently of observation, and
contrary to David Mermin's statement\cite{Mermin1981b}, I would thus
argue that Einstein's attacks against the metaphysical underpinning
of quantum theory are still valid today, and do not contradict
nature itself.

\bibliographystyle{unsrt} 
\bibliography{refmoon}

\end{document}